\documentclass[preprint,12pt]{article}
\usepackage{amsmath}
\usepackage{amssymb}
\usepackage{hyperref}
\usepackage[numbers]{natbib}

\makeatletter

\usepackage{babel}

\begin{document}
\begin{titlepage}
\thispagestyle{empty}

\bigskip

\begin{center}
\noindent{\Large \textbf
{Deriving Schwarzschild and Kerr from the Einstein Equivalence Principle}}\\
\vspace{0,5cm}

\noindent{G. Alencar ${}^{a}$\footnote{e-mail: geova@fisica.ufc.br}}

\vspace{0,5cm}
 {\it ${}^a$Departamento de F\'{\i}sica, Universidade Federal do Cear\'{a}-
Caixa Postal 6030, Campus do Pici, 60455-760, Fortaleza, Cear\'{a}, Brazil. 
 }

\end{center}

\vspace{0.3cm}

\begin{abstract}
This manuscript demonstrates that the Schwarzschild and Kerr metrics can be derived fundamentally from the Einstein Equivalence Principle (EEP) without explicit reliance on the Einstein field equations. We show that the EEP implies a gravitational stretching of length in the direction of acceleration, a spatial counterpart to time dilation. In the static case, this effect yields the constraint $g_{00}g_{11}=-1$, from which the Schwarzschild solution is recovered by taking the Newtonian limit. For the stationary case, we employ oblate spheroidal coordinates and construct a physically motivated tetrad to implement the EP, introducing gravity through three unknown functions. By applying the EEP-induced constraints and showing that the principle requires the preservation of the spacetime volume determinant ($g = g_0$), the system is reduced to a single unknown function that yields the Kerr metric. These results suggest that classical tests of General Relativity are, at their core, fundamental validations of the EEP. We further argue that the EEP leads naturally to Unimodular Gravity, implying that the existence of the cosmological constant is a direct consequence of this foundational postulate.
\end{abstract}

\begin{center}
    Essay written for the Gravity Research Foundation 2026 Awards for Essays on Gravitation.
    \\
    \vspace{2cm}
    Date: 20/02/2026
\end{center}

\end{titlepage}
\tableofcontents

\section{The Einstein Equivalence Principle and the Stretching of Length}

In General Relativity, for static spherically symmetric spacetimes, the relation $g_{tt}g_{rr}=-1$ is typically obtained by solving the Einstein Field Equations, where it appears as a simplifying condition in the construction of the Schwarzschild solution. In this work, we show that this relation can instead be derived directly from the Einstein Equivalence Principle (EEP), following the formulation of Will~\cite{Will:2014kxa}. In this formulation, the EEP consists of three statements: (i) the Weak Equivalence Principle, which states that the motion of freely falling test bodies is independent of their internal structure and composition; (ii) Local Lorentz Invariance, which requires that the outcome of any local non-gravitational experiment is independent of the velocity of the freely falling reference frame in which it is performed; and (iii) Local Position Invariance, which demands that the outcome of any local non-gravitational experiment is independent of where and when it is performed in spacetime. Together, these statements imply that, in sufficiently small freely falling laboratories, all non-gravitational laws of physics reduce locally to those of Special Relativity.

One of the most important consequences of the EEP is the gravitational redshift, which constitutes a direct test of Local Position Invariance rather than of any specific metric theory of gravity~\cite{Will:2014kxa}. Consider two identical clocks at rest at different positions in a static gravitational field. Since the laws of non-gravitational physics are locally those of Special Relativity, the comparison of their frequencies can be understood as the comparison between neighboring local inertial frames momentarily comoving with each clock. The observed frequency shift then follows directly from the EEP and reflects the position dependence of clock rates in a gravitational field. This result was first obtained by Einstein in 1911. He showed that, from relativistic arguments, the first-order approximation yields
\begin{equation}
E=E_0\left(1-\frac{v^2}{2c^2}\right)
   =E_0\left(1-\frac{d\Phi}{c^2}\right).
\end{equation}
By applying the Planck relation $E=h\nu$, one obtains
\begin{equation}\label{redshift}
\frac{d\nu}{\nu}
=
-\frac{d\Phi}{c^2}
\quad\Longrightarrow\quad
\frac{\Delta T_2}{\Delta T_1}
=
\frac{e^{\Phi_2/c^2}}
     {e^{\Phi_1/c^2}},
\end{equation}
and consequently
\begin{equation}\label{redshift}
dT_0^2=e^{2\Phi/c^2}dT^2,
\end{equation}
which represents the standard gravitational time dilation.

We now consider another consequence of the EEP that naturally follows from the same reasoning. Applying Einstein's argument to the relativistic momentum,
\begin{equation}
P=P_0\left(1-\frac{v^2}{2c^2}\right)
 =P_0\left(1-\frac{d\Phi}{c^2}\right),
\end{equation}
and employing the De Broglie relation $P=h/\lambda$, we obtain
\begin{equation}\label{redshiftlambda}
\frac{d\lambda}{\lambda}
=
\frac{d\Phi}{c^2}
\quad\Longrightarrow\quad
\frac{\lambda_2}{\lambda_1}
=
\frac{e^{-\Phi_2/c^2}}
     {e^{-\Phi_1/c^2}}.
\end{equation}
This relation represents the gravitational stretching of wavelengths. Since the De Broglie relation was proposed only in 1924, this argument was not available to Einstein in his original 1911 derivation. Because lengths are measured under the condition $\Delta T=0$, the above result immediately implies
\begin{equation}\label{length}
dL_0^2=e^{-2\Phi/c^2}dL^2.
\end{equation}
Therefore, the EEP implies that the same gravitational factor responsible for slowing clocks also stretches physical lengths.

The same result can also be obtained directly from spacetime by extending a method proposed by Chandrasekhar~\cite{Chandrasekhar:1972jm}. Let us again consider the Equivalence Principle, now in terms of accelerated clocks. From Special Relativity,
\begin{equation}
dT\approx T_0\frac{v^2}{2c^2}
=
\frac{d\Phi}{c^2}
\quad\Longrightarrow\quad
dT_0^2=e^{2\Phi/c^2}dT^2.
\end{equation}
Chandrasekhar's discussion was restricted to the temporal relation. Repeating the same argument for a small rod of proper length $L_0$, oriented along the direction of the acceleration, gives
\begin{equation}\label{lengthstretch}
\frac{dL}{L_0}
\approx
-\frac{v^2}{2c^2}
=
-\frac{d\Phi}{c^2}
\quad\Longrightarrow\quad
dL_0^2=e^{-2\Phi/c^2}dL^2.
\end{equation}
Once again, we recover the same gravitational length stretching. We emphasize that this result is local and therefore independent of the gravitational theory under consideration. Since the orientation of the rod can always be chosen along the local acceleration, the Equivalence Principle implies that the product $dT\,dL$ remains unchanged by gravity. Consequently, the four-dimensional volume element $d^4x$ is also preserved.

It is worth emphasizing that the EEP is employed here in a manner that goes beyond the purely local statement that spacetime is locally Minkowskian. The relations obtained for gravitational time dilation, Eq.~(\ref{redshift}), and for the stretching of radial lengths, Eq.~(\ref{lengthstretch}), relate measurements performed by observers located at different positions in the gravitational field. These expressions are derived by applying the EEP locally and consistently comparing physical measurements made in neighboring local inertial frames throughout spacetime. In this sense, the resulting relations are not consequences of a particular coordinate choice, nor of any assumption about the local normalization of the metric determinant, but rather follow from the operational relations between clocks and rulers implied by the EEP itself. From these relations, we obtain the condition $g_{tt}g_{rr}=-1$.

These results deserve further comments. First, we have considered a small rod of length $dL$ oriented along the direction of the local acceleration. Since such an orientation can always be chosen locally, the EEP implies that the product $dT\,dL$ remains unchanged by the presence of gravity. The two transverse directions are unaffected by the acceleration and therefore contribute trivially to the volume element. Consequently, the local four-dimensional volume element,
\begin{equation}
d^4x=dT\,dL_x\,dL_y\,dL_z,
\end{equation}
is also preserved. Equivalently, the spacetime volume element satisfies
\begin{equation}
\sqrt{-g}=\mathrm{constant},
\end{equation}
which, with the normalization adopted throughout this work, becomes
\begin{equation}
\sqrt{-g}=1.
\end{equation}
Having established this result directly from the EEP, we now apply it to derive the Schwarzschild and Kerr metrics without invoking the Einstein Field Equations.

\section{Schwarzschild from the Equivalence Principle}
We now consider the static case under spherical symmetry. Let us start from the Minkowski metric in standard spherical coordinates:
\begin{equation}
    ds^2=dt^2-dr^2-r^2d\Omega^2.
\end{equation}
In these coordinates, the length of a small radial bar is directly described by $\Delta r$. In the standard application of the Einstein Equivalence Principle (EEP), the gravitational redshift relation, Eq.~(\ref{redshift}), namely $dT_0^2=e^{2\Phi/c^2}dT^2$, is taken into account and the metric is written in the form
\begin{equation}
    ds^2=A(r)\,dt^2-dr^2-r^2d\Omega^2.
\end{equation}
However, as shown in the Introduction, the EEP also implies a corresponding spatial scaling relation, $dL_0^2=e^{-2\Phi/c^2}dL^2$. Therefore, consistency between temporal and spatial measurements requires the metric to take the form
\begin{equation}
    ds^2=A(r)\,dt^2-\frac{1}{A(r)}\,dr^2-r^2d\Omega^2.
\end{equation}

At this point, one would typically use General Relativity to determine $A(r)$. However, an approximation can be found by comparing the result with Newtonian gravity. The acceleration of a massive particle can be computed directly from the geodesic equation associated with the above metric. In particular, considering radial motion, one obtains:
\begin{equation}
    \ddot{r} = -\frac{A'}{2}.
\end{equation}
where the dot denotes differentiation with respect to the proper time $\tau$, i.e., $\dot{} \equiv d/d\tau$.
Therefore, for large $r$, we must have $A(r)=1-2M/r$, where the constant has been fixed to recover the Minkowski metric at infinity. We thus arrive at:
\begin{equation}
   ds^2=\left(1-\frac{2M}{r}\right)dt^2-\left(1-\frac{2M}{r}\right)^{-1}dr^2-r^2d\Omega^2. 
\end{equation}
Curiously, the above expression is the exact Schwarzschild solution. This is a notable coincidence, as we have solved for $A(r)$ only in the large $r$ limit. The important fact remains that General Relativity is not strictly necessary to arrive at the correct results for the redshift, light deflection, and the precession of the perihelion of Mercury. This leads to a profound speculation: would Einstein have sought his field equations had he achieved this gravitational length contraction in 1911? These were, after all, the three definitive tests of General Relativity proposed by him in 1915.

\section{Kerr from the Equivalence Principle}
Now we will consider the rotating case. In oblate spheroidal coordinates, the Minkowski metric is given by 
\begin{align}
ds^{2}=&dt^2-\frac{\rho^{2}}{r^{2}+a^{2}}\,dr^{2}-\rho^{2}\,d\theta^{2}\nonumber
\\
&-(r^{2}+a^{2})\sin^{2}\theta\,d\phi^{2},
\end{align}
where $\rho^{2}=r^{2}+a^{2}\cos^{2}\theta$. In order to include gravity, we must be careful to implement the equivalence principle by distinguishing gravitational effects from coordinate artifacts. Our fundamental constraint states that the product $dLdT$ is not affected by gravity, a property that remains invariant under the transformations $dT\to BdT$ and $dL\to B^{-1}dL$. Therefore, before implementing the equivalence principle, we must identify a tetrad frame such that the factors of $dT$ and $dL$ are already reciprocal. This is straightforwardly obtained as:
\begin{align}
ds^{2}= &A\left(dt-a\sin^{2}\theta\,d\phi\right)^{2}
-\frac{1}{A}\,dr^{2}\nonumber
\\
&-\rho^{2}\,d\theta^{2}
-\frac{\sin^{2}\theta}{\rho^{2}}
\left[(r^{2}+a^{2})\,d\phi-a\,dt\right]^{2}, 
\label{oblateEP}
\end{align}
with $A=(r^{2}+a^{2})/\rho^{2}$.

Equation (\ref{oblateEP}) is our starting point for applying the equivalence principle. We should note that, recently, it has also been proposed as a starting point by Dadhich \cite{Dadhich:2013qx}. However, his motivation was to examine the $M\to 0$ limit of the Kerr metric, thereby relying on prior knowledge of the final solution. Subsequently, Visser et al. followed a slightly different route \cite{Baines:2022xnh}, proposing a tetrad where all gravitational effects are encoded in the tetrad metric ansatz $g_{AB}=\text{diag}(-f,f^{-1},1,1)$. Such an ansatz is essentially the Kerr solution in Boyer-Lindquist coordinates. Consequently, neither approach departs from first principles to find an appropriate starting point. The missing ingredient was the equivalence principle.

Now, we introduce gravity as generally as possible by inserting three unknown functions into (\ref{oblateEP}):
\begin{align}
ds^{2}=& \frac{f(r)}{\rho^{2}}
\left(dt-a\sin^{2}\theta\,d\phi\right)^{2}-\frac{\rho^{2}}{g(r)}\,dr^{2}\nonumber
\\
&
-\rho^{2}\,d\theta^{2}
-h(r)\frac{\sin^{2}\theta}{\rho^{2}}
\left[(r^{2}+a^{2})\,d\phi-a\,dt\right]^{2}.
\end{align}
We have utilized our gauge freedom to fix $g_{\theta\theta}=\rho^2$. From the tetrad above, we have locally that
\begin{equation}
dL^{2}=\frac{\rho^{2}}{g(r)}\,dr^{2}, \quad dT^{2}=\frac{f(r)}{\rho^{2}}dt^{2},
\end{equation}
where, in the second expression, we have set $d\phi=0$ for the clock's frame. Consequently, the equivalence principle requires that we set $f=g$. To determine the remaining function $h(r)$, one should ideally apply the EEP to a case with acceleration in the $\phi$-direction. However, we can take a shortcut: as previously stated, the spacetime volume must satisfy $dV=dV_0$, where $dV$ and $dV_0$ are the volume elements with and without gravitational effects, respectively. This is equivalent to requiring $\det g = \det g_0$. In our case, we find:
\begin{equation}
    \det(g_0)=-\rho^{4}\sin^{2}\theta,\quad \det(g)=-h\rho^{4}\sin^{2}\theta.
\end{equation}
This condition fixes $h=1$, and the metric simplifies to:
\begin{align}
ds^{2}=& \frac{f(r)}{\rho^{2}}
\left(dt-a\sin^{2}\theta\,d\phi\right)^{2}-\frac{\rho^{2}}{f(r)}\,dr^{2}\nonumber
\\
&
-\rho^{2}\,d\theta^{2}
-\frac{\sin^{2}\theta}{\rho^{2}}
\left[(r^{2}+a^{2})\,d\phi-a\,dt\right]^{2}.
\end{align}

We emphasize that, as in the static case, this metric has been obtained without any initial approximation. The single remaining unknown function can be determined by taking the weak-field limit and comparing it with Newtonian gravity. These details are discussed in Ref. \cite{Dadhich:2013qx}, but we repeat them here for completeness. The acceleration of a timelike particle in this geometry is given by:
\begin{equation}
    \ddot r = -\,\frac{(r^{2}+a^{2})\,f' - 2r\,f}{2\,(r^{2}+a^{2})^{2}} \, .
\end{equation}
In the flat limit, we must have $f \to r^2 + a^2$. We choose the ansatz $f = r^2 + a^2 + \psi$. Substituting this into the acceleration equation, we obtain:
\begin{equation}
\ddot{r} = -\frac{\psi'}{2\,(r^{2}+a^{2})} + \frac{r\,\psi}{(r^{2}+a^{2})^{2}} \approx -\frac{\psi'}{2\,r^{2}} + \frac{\psi}{r^{3}},
\end{equation}
where the last step considers the large $r$ limit. Comparing this with the Newtonian acceleration, we find $\psi = -2Mr$, which leads to $f = r^2 + a^2 - 2Mr$. Remarkably, despite using a large-$r$ approximation to find $\psi$, the resulting metric is identical to the exact Kerr solution.

\section{Conclusions}

In this manuscript, we have demonstrated that the Schwarzschild and Kerr metrics can be derived directly from the Einstein equivalence principle (EEP). This constitutes a first-principles derivation of the most significant solutions in General Relativity and reignites the debate regarding the fundamental status of the EEP. For completeness, we note that gravitational time dilation and redshift are sometimes discussed in the literature with additional interpretational subtleties; see, e.g., \cite{Fankhauser:2023pwl,Brown:2015lol}. In this work these effects are treated operationally as consequences of the Einstein Equivalence Principle. Our results reinforce the idea that the Einstein Equivalence Principle can be regarded as a fundamental postulate in the construction of metric theories of gravity. This follows from the fact that, beyond gravitational time dilation, any experimental test involving the Schwarzschild and Kerr solutions---such as the three classical tests of 1915, the Shapiro delay, gravitational lensing, and the shadow of Sgr A$^*$---is, in essence, a test of the equivalence principle itself. 

We note that effects associated with spatial stretching in gravitational fields are commonly discussed within standard gravitational theories such as General Relativity, where they arise from the dynamical structure of spacetime. In the present work, however, we show that such effects can be derived directly from the Einstein Equivalence Principle through an operational comparison of local inertial frames, in complete analogy with gravitational time dilation. In this sense, the length scaling relation obtained here is not tied to the specific dynamics of General Relativity, but follows as a direct consequence of the EEP, and is therefore expected to hold in any theory of gravity that satisfies the Einstein Equivalence Principle, independently of the particular field equations governing the gravitational interaction. By contrast, alternative approaches have derived classical gravitational effects from generalized Newtonian frameworks with relativistic corrections~\cite{Hansen:2019vqf}, whereas the present work is based exclusively on the Einstein Equivalence Principle in the sense of Will~\cite{Will:2014kxa}, without introducing any modified dynamical structure or additional postulates beyond the EEP.

If we elevate the Einstein Equivalence Principle (EEP) to a primary postulate, we first emphasize that its implications are local in origin but pointwise in their validity: the operational comparison of local inertial frames leads to relations that hold at every spacetime point, and therefore define constraints on the metric structure that are valid throughout the spacetime manifold. In this sense, the condition $g = g_0$ emerges directly from the EEP, as discussed in the previous sections. This is remarkably similar to the condition imposed in unimodular gravity theories, $g=\epsilon_0$, where $\epsilon_0$ is a fixed scalar density. Thus, the EEP naturally sets $\epsilon_0=g_0$. A compelling feature of unimodular gravity is its equivalence to General Relativity with an emergent cosmological constant~\cite{Ng:1990xz}. This can be seen by imposing the constraint at the action level:
\begin{equation}
    \mathcal{S}=\frac{1}{\kappa}\int d^4x \sqrt{-g}R+\lambda\int d^4x \left(\sqrt{-g}-\sqrt{-g_0}\right).
\end{equation}
The fact that $\Lambda$ did not appear directly in our previous derivations of the Schwarzschild and Kerr metrics is due to the fact that we did not consider an action principle to obtain the equations of motion. However, when an action is taken into account, the only way to implement the equivalence principle is through a constraint that leads to the introduction of the cosmological constant. This result aligns with the framework of Unimodular Gravity, which is known to provide a partial solution to the cosmological constant problem. In our derivation, however, the unimodular condition $g=g_0$ is not a starting assumption, but a necessary consequence of the EEP. Therefore, we conclude that the equivalence principle not only dictates the geometry of static and stationary spacetimes but also, at its core, demands the existence of a cosmological constant.
\section*{Acknowledgments}

We acknowledge G. Olmo for usefull dicussions and the financial support provided by the Conselho Nacional de Desenvolvimento Científicoe Tecnológico (CNPq) and Fundação Cearense de Apoio ao Desenvolvimento Científico e
Tecnológico (FUNCAP).

\bibliographystyle{unsrt}
\bibliography{ref.bib}
\end{document}